\documentclass[letterpaper, 10 pt, conference]{IEEEtran}

\usepackage[noadjust]{cite}
\usepackage{graphicx,xcolor,epstopdf}
\usepackage{amsmath,amsfonts,amssymb,amsthm,mathtools}
\usepackage{stfloats}
\usepackage{booktabs}
\usepackage{tikz,tikzinclude}
\usetikzlibrary{intersections}
\usetikzlibrary{shapes,arrows,calc}
\usetikzlibrary{chains,positioning,quotes,patterns}

\usepackage{hyperref}
\hypersetup{
	colorlinks,
	linkcolor={violet!75!black},
	citecolor={green!50!black},
	urlcolor={blue!50!black}
}

\tikzset{BlockDiagram/.style={
                block/.style = {draw, rectangle, rounded corners, text centered, minimum height=1cm, minimum width=1cm},
                gain/.style = {draw, isosceles triangle, text centered, minimum height=1cm, minimum width=1cm},
                sum/.style = {draw, circle, thick, minimum size=4mm, node distance=5mm, inner sep=0mm},
                input/.style = {coordinate},
                point/.style = {coordinate},
                output/.style = {coordinate},
}} 

\newtheorem{theorem}{Theorem}
\newtheorem{prop}{Proposition}

\newtheorem{definition}{Definition}
\theoremstyle{remark}
\newtheorem*{remark}{Remark}

\newcommand{\R}[1]{\mathbb{R}^#1}
\newcommand{\Lie}[1]{\mathcal{L}_{#1}}

\begin{document}

\title{\textit{Inverse Optimal Control Design of a VTOL Aircraft}}

\author{\IEEEauthorblockN{Kinza Rehman\IEEEauthorrefmark{1}, Hafiz Zeeshan Iqbal Khan\IEEEauthorrefmark{2}\IEEEauthorrefmark{3}, and Muhammad Farooq Haydar\IEEEauthorrefmark{3}\vspace{0.15cm}}
\IEEEauthorblockA{\IEEEauthorrefmark{1} School of Mechanical and Manufacturing Engineering, National University of Sciences and Technology, Islamabad.}
\IEEEauthorblockA{\IEEEauthorrefmark{2} Centers of Excellence in Science and Applied Technologies, Islamabad.}
\IEEEauthorblockA{\IEEEauthorrefmark{3} Department of Aeronautics \& Astronautics, Institute of Space Technology, Islamabad.}}

\maketitle

\begin{abstract}
\textit{\mdseries A Vertical Takeoff and Landing (VTOL) aircraft is capable of short/vertical takeoffs and landings, thus eliminating the requirement of long runways. This feature makes them suitable for a broader variety of missions, generally not achievable by a traditional aircraft. The dynamics of the VTOL aircraft in the vertical plane has been used as a benchmark problem to demonstrate the effectiveness of different nonlinear control techniques, mainly due to being highly coupled and non-minimum phase. This paper presents the design of an inverse optimal control for a simplified VTOL aircraft model. The proposed control law is based on a Control Lyapunov Function (CLF) which exploits the normal form typically used for control design through feedback linearization. The suggested control law is compared with dynamic feedback linearization based control law and a comparison is drawn based on their efficacy and robustness.\\}
\end{abstract}

\begin{IEEEkeywords}
\textit{Inverse Optimal Control; VTOL Aircraft; Control Lyapunov Function; Dynamic Feedback Linearization}
\end{IEEEkeywords}

\section{Introduction}

Vertical Takeoff and Landing (VTOL) aircraft are distinguished from conventional aircraft, as the name implies, prominently because of their ability to take off and land vertically. Thus eliminating the requirement of long stretches of runways \cite{Zhang2017}. Due to this feature, they can be used in a much wider range of missions, as compared to traditional airplanes. Such aircraft have gained attention worldwide due to their application in complex operations that further improve their combat effectiveness, making them an excellent choice for defense purposes. World’s first jet VTOL, the British “Harrier” fighter \cite{Intwala2015} and the Jacques series fighter by the Soviet Union are among the aircraft relying mainly on jet thrust vector technology. Multiple types of technologies are being used by different VTOL designs. A type of VTOL aircraft can change its thrust direction by using nozzles either vertically or horizontally while having fixed engines. However, this process can bring back the high-temperature exhaust towards the aircraft, consequently decreasing engine efficiency and may cause damage to the surface of the aircraft. Another type of VTOL technology is tilted rotorcraft, the rotor shaft is perpendicular to the ground during takeoff and landing and an upward lifting force is provided by the propeller. A few examples of such aircraft are Osprey V-22, XV-3, and XV-15.

Despite of immense application value and development prospects, the control of a VTOL aircraft possesses significant challenges due to their highly coupled, nonlinear, and non-minimum phase dynamics \cite{Sastry1991}. As a result, it has grabbed the attention of the control community, and it is frequently used as a benchmark problem for evaluating novel control strategies. Many linear and nonlinear control techniques \cite{Sanahuja2007} have been applied to this problem, and many different approaches have been considered including static state feedback \cite{OlfatiSaber2002}, output-feedback \cite{Do2003}, as well as tracking of time-varying trajectories \cite{Ailon2010}.

Optimal control theory determines the physical constraint satisfying the control signal and minimizes the performance cost function to attain a certain optimality criterion simultaneously \cite{Kirk2004}. Though the problem of optimal control of linear systems is pretty much solved, the optimal control of nonlinear systems \cite{Chen2003} still poses a challenge, as it requires solving Hamilton–Jacobi–Bellman (HJB) partial differential equations (PDEs). Which for most practical scenarios are not analytically solvable, therefore solution is generally obtained through numerical approaches \cite{Polak1997}, which in turn usually requires a lot of computational resources. Many desirable features of a closed-loop system are guaranteed by optimal stabilization, such as stability margins.
And, in general, these robustness characteristics are independent of the cost functional used. Freeman and Kokotovi\'{c} \cite{Freeman1995} pursued the development of inverse optimal control techniques due to these beneficial characteristics. Inverse optimal control, rather than optimizing a given cost function, computes the cost function being optimized by a given stabilizing control law for a known system dynamics. Kalman \cite{Kalman1964} first introduced inverse optimal control by defining the case of autonomous LQR. He took a single-input case and found conditions (both necessary and sufficient) for a feedback controller to be optimal. Later on, further working was done on nonlinear problems, and multiple solutions for different classes of systems are proposed \cite{Thau1967, Moylan1973, Krstic1998, Ng2000, Johnson2013, Pauwels2014, Jean2018}. Inverse optimal control finds its applications in multiple domains ranging from human physiological movements \cite{Todorov2002} to goal-oriented human locomotion \cite{Mombaur2010, Chitour2010}. A similar technique is applied on animal-inspired millimeter-scaled micro quadrotors in \cite{Faruque2018}. In robotics, similar control is applied for reliable and comfortable autonomous robotic cars \cite{Kuderer2015}, whereas, in social sciences, inverse optimal control helps in finding utility functions for decision-makers \cite{Nielsen2004}.

In this paper, an inverse optimal control law for a VTOL aircraft is designed. Due to strong input coupling and non-minimum phase dynamics, the system is transformed to obtain a dynamic feedback linearization based control law and corresponding normal form. The normal form is used in constructing Control Lyapunov Function (CLF), leading to the design of inverse optimal control law. To demonstrate the efficacy and robustness, the inverse optimal control law is compared with dynamic feedback linearization based controller, in both nominal and perturbed scenarios. The rest of the paper is organized as follows: In section \ref{sec:Prelim}, the preliminaries are discussed, followed by the dynamics of \emph{Harrier Jump Jet} aircraft in section \ref{sec:VTOLDyn}. The inverse optimal control design is discussed in section \ref{sec:InvOPT}, while section \ref{sec:Res} contains simulation results and discussion, and section \ref{sec:Conc} concludes the publication. 

\section{Preliminaries}\label{sec:Prelim}
The fundamental obstacle in solving an optimal control problem is the HJB equation solution, which frequently overshadows  numerous desired aspects of optimal stabilization for closed-loop systems; this led to the development of the inverse problem of optimal stabilization \cite{Kokotovic2012}. The main difference between the two approaches is in the latter first step is the design of stabilizing feedback and the second is to find a cost function for which it is optimal.

This section will cover the fundamentals of the inverse optimal method after presenting some significant results from the literature regarding optimal controllers and their stability guarantees. \emph{Hamilton-Jacobi-Bellman} (HJB) equations and \emph{Pontryagin Maximum Principle} (PMP) are two types of optimality conditions. However, the former is generally appropriate for feedback control design across infinite horizons \cite{Kokotovic2012}. One of the key benefits of an optimal or inverse optimal controller, which gives them strong robustness margin guarantees, is dominating the undesired nonlinearities, instead of cancelling them, as in feedback linearization.

\subsection{Optimal Control}
Consider a nonlinear system of form
\begin{equation}\label{eq:Prelim01}
  \dot{x} = f(x) + g(x)u
\end{equation}
where $x\in\R{n}$ and $u\in\R{m}$ are states and inputs, respectively. Also for any scalar function $h:\R{n}\mapsto\mathbb{R}$, we denote its Lie derivative along a vector fields $f$ as $\Lie{f}h \triangleq \left(\frac{\partial h}{\partial x}\right)^\top f$.

\begin{definition}[Optimal Feedback Control]
A control law $u^*(x)$ is optimal if it renders the equilibrium of \eqref{eq:Prelim01} to be asymptotically stable, and it minimizes the cost functional
\begin{equation}\label{eq:Prelim02}
  J = \int_{0}^{\infty}\big[ \mathcal{Q}(x) + u^\top \mathcal{R}(x) u \big]\,\mathrm{d}t
\end{equation}
where $\mathcal{Q}(x)\ge0$ and $\mathcal{R}(x)>0$ for all $x$.
\end{definition}

\begin{theorem}[Optimal Stabilization {\cite[Theorem 3.19]{Kokotovic2012}}]\label{thm:01}
Assume that the HJB equation \eqref{eq:Prelim03} is satisfied by a $\mathcal{C}^1$ positive semidefinite function $V(x)$
\begin{equation}\label{eq:Prelim03}
 \mathcal{Q}(x) + \Lie{f}V(x) - \frac{1}{4}[\Lie{g}V(x)]\mathcal{R}^{-1}(x)[\Lie{g}V(x)]^\top = 0
\end{equation}
where $V(0) = 0$, and such that the feedback control law
\begin{equation}\label{eq:Prelim04}
 u^*(x) = -\frac{1}{2}\mathcal{R}^{-1}(x)[\Lie{g}V(x)]^\top
\end{equation}
guarantees asymptotic stability of the origin. Then $u^*(x)$ becomes optimal stabilizing control law which minimizes the cost \eqref{eq:Prelim02} over all $u$ while assuring  $\lim_{t\rightarrow\infty} x(t) = 0$, and $V(x)$ be the optimal value function.
\end{theorem}

\begin{figure}
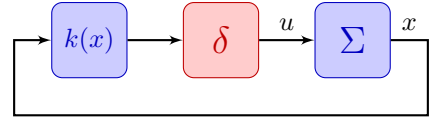

  \centering
  \includetikzgraphics[BlkDiag]{FiguresTikz.tex}
  \caption{Nonlinear Stability Margins: Feedback Interconnection}
  \label{fig:BlkDiag}
\end{figure}
\begin{definition}[Stability Margins]
The nominal ($\delta = I$) nonlinear feedback interconnection $(\Sigma,k)$ (see Fig. \ref{fig:BlkDiag})  is said to have a stability (gain, sector or disk) margin given the perturbed closed-loop system $(\Sigma,k,\delta)$ (see Fig. \ref{fig:BlkDiag})  is GAS for all $\delta\in\Delta$, where $\Delta$ is of the following forms:
\begin{itemize}
  \item Gain margin $(\alpha,\beta)$: $\Delta = \mathrm{diag}(\kappa_1,\cdots,\kappa_m)$ with constants $\kappa_i \in (\alpha,\beta)\,\forall\,i \in[1,m]$
  \item Sector margin $(\alpha,\beta)$: $\Delta = \mathrm{diag}(\psi_1(\cdot),\cdots,\psi_m(\cdot))$ where $\psi_i(\cdot)$'s are locally Lipschitz static nonlinearities belonging to sector $(\alpha,\beta)$.
  \item Disk margin $D(\alpha)$ : $\Delta$ is GAS and IFP($v$) (Input Feedforward Passive), $ v > \alpha$, with a storage function having radial unboundedness.
\end{itemize}
\end{definition}

\begin{prop}\label{prop:01}
A feedback interconnection of system \eqref{eq:Prelim01} and optimal stabilizing controller \eqref{eq:Prelim04}, which minimizes the cost functional \eqref{eq:Prelim02}, has
\begin{itemize}
  \item Disk margin $D\left(\frac{1}{2}\right)$ if $\mathcal{R}(x) = I$  \cite[Proposition 3.31]{Kokotovic2012}
  \item Sector margin $\left(\frac{1}{2},\infty\right)$ if $\mathcal{R}(x)$ is diagonal \cite[Proposition 3.34]{Kokotovic2012}
\end{itemize}
\end{prop}

\subsection{Inverse Optimal Control}

\begin{theorem}[Inverse Optimal Stabilization \cite{Kokotovic2012}]\label{thm:02}
A control law $u^*(x)$ is inverse optimal (globally), and minimizes the cost functional \eqref{eq:Prelim02}, if it renders the equilibrium of \eqref{eq:Prelim01} to be (globally) asymptotically stable, and can be of the form
\begin{equation}\label{eq:Prelim05}
  u^*(x) = -\frac{1}{2}\mathcal{R}^{-1}(x)\Lie{g}V(x)
\end{equation}
where $\mathcal{R}(x)>0$, and $V(x)>0$ is (radially unbounded) positive semidefinite function whose derivative is negative semidefinite with the control $u(x) = u^*(x)/2$, i.e. $\Lie{f}V + \frac{1}{2}\Lie{g}Vu^* < 0$. And the corresponding $\mathcal{Q}(x)$ in cost functional \eqref{eq:Prelim02} can be computed as
\begin{equation}\label{eq:Prelim06}
  \mathcal{Q}(x) = -\Lie{f}V(x) -\frac{1}{2} \Lie{g}V(x)u^*(x)
\end{equation}
\end{theorem}

Instead of cancelling nonlinearities, the inverse optimal approach presents a constructive alternative to achieve desired stability margins and optimality properties using Lyapunov function construction.
The Lyapunov function chosen not only focuses on features of $f(x)$, but it also takes into account the freedom given by the term $g(x)u$. For a generic nonlinear system \eqref{eq:Prelim01}, Artstein and Sontag \cite{Sontag1989} introduced CLF, defined below.
\begin{definition}[Control Lyapunov Function]
Given a nonlinear system \eqref{eq:Prelim01}, a $\mathcal{C}^1$ positive definite, and radially unbounded function $V(x)$ is said to be \emph{Control Lyapunov Function} (CLF) if $\Lie{g}V(x) = 0$ implies $\Lie{f}V(x) < 0$ for all $x\ne0$.
\end{definition}

There could be many ways to construct a CLF, one approach could be to transform the system \eqref{eq:Prelim01} into its normal form, assuming it is input-state feedback linearizable, then because the normal form is transformable to a linear system via feedback, constructing a CLF for the linear system can provide a CLF for the nonlinear system. For a known CLF, considering it as an optimum value function may assist in picking the inverse optimal stabilizing control law from a set of explicit expressions \cite{Freeman1995}. Sontag's formula is an example of an optimum control law \cite{Sontag1989}

\begin{equation}\label{eq:Prelim07}
u_{S} \triangleq \begin{cases} - \left[c_0 + \dfrac{\alpha + \sqrt{\alpha^2 + (\beta^\top\beta)^2}}{\beta^\top\beta}\right]\beta, &\quad \beta \ne 0 \\
         \hfil 0,  &\quad \beta = 0
   \end{cases}
\end{equation}
where $c_0>0$ is constant and
\[
\alpha(x) \triangleq \Lie{f}V(x), \qquad \beta(x) \triangleq (\Lie{g}V(x))^\top
\]

\begin{prop}[Optimal Cost of Sontag's Formula {\cite[Proposition 3.44]{Kokotovic2012}}]\label{prop:02}
The control law \eqref{eq:Prelim07} is inverse optimal with the cost functional \eqref{eq:Prelim02}, where
\begin{equation}\label{eq:Prelim08}
\begin{split}
  \mathcal{Q}(x) &= \frac{1}{2}\mu(x)\beta(x)^\top\beta(x),\\
  \mathcal{R}(x) &= \frac{1}{2}\mu(x)I_2
\end{split}
\end{equation}
where,
\[
\mu(x) = \begin{cases} c_0 + \dfrac{\alpha + \sqrt{\alpha^2 + (\beta^\top\beta)^2}}{\beta^\top\beta}, &\quad \beta \ne 0 \\
         \hfil c_0,  &\quad \beta = 0
   \end{cases}
\]
\end{prop}

\begin{remark}
It is worth noting that, since $\mathcal{R}(x)$ is diagonal and not constant, so from Proposition \ref{prop:01} we can guarantee the control law \eqref{eq:Prelim07} has a sector margin of $(\frac{1}{2},\infty)$.
\end{remark}

\section{Harrier Jump Jet Aircraft Dynamics}\label{sec:VTOLDyn}

\emph{Harrier} ``jump jet'', a vectored thrust aircraft, is manufactured by McDonnell Douglas (Boeing) and the model name is \emph{YAV-8B Harrier}. It is a light attack, single-seat transonic aircraft. A single Turbofan engine is responsible for powering the aircraft, with its four rotatable exhaust nozzles. These nozzles can be rotated from their normal aft position to 100 degrees forward, which helps in nozzle braking. The aircraft has a reaction control system (RCS), responsible for providing moment generation, complementary to its traditional aerodynamic control surfaces. The \emph{Harrier} achieves its capability of vertical and short takeoff and landing (V/STOL) by redirecting its thrust downward and using small maneuvering thrusters in RCS located on its wings. To study VTOL dynamics, it is generally convenient to consider aircraft motion restricted to the vertical plane only, commonly known as planar vertical takeoff and landing (PVTOL) dynamics.

\subsection{PVTOL Dynamics}
\begin{figure}
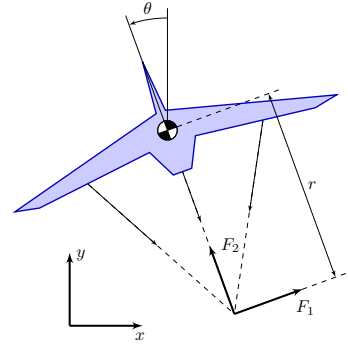

  \centering
  \scalebox{0.65}{\includetikzgraphics[PVTOL]{FiguresTikz.tex}}
  \caption{PTVOL Aircraft: An Schematic (Adapted from \cite{Astrom2012})} \label{fig:01}
\end{figure}

Figure \ref{fig:01} depicts a simplified Harrier model, which only considers vehicle motion in the vertical plane \cite{Astrom2012}. The net thrust produced by a primary thruster facing downwards and the maneuverable ones may be expressed as two forces $F_1$ and $F_2$ having distance $r$.
Let ($x,y,\theta$) be the position and orientation of the aircraft's center of mass. The equations of motion are expressed as,
\begin{equation}\label{eq:AstromPVTOL}
\begin{split}
m\ddot{x} &= F_1\cos\theta - F_2\sin\theta \\
m\ddot{y} &= F_1\sin\theta + F_2\cos\theta - mg\\
J\ddot{\theta} &= r F_1
\end{split}
\end{equation}
where, $m$ is vehicle's mass, $J$ represents moment of inertia, $g$ is acceleration due to gravity. Here it must be noted that the aerodynamic damping has been ignored, which is generally quite negligible at slow speeds during vertical takeoff and landing. For better analysis and comparison, we consider the following normalized form of \eqref{eq:AstromPVTOL}.
\begin{equation}\label{eq:PVTOL}
\begin{split}
   \ddot{x}    &= \epsilon \tau \cos\theta - f \sin\theta \\
   \ddot{y}    &= \epsilon \tau \sin\theta + f \cos\theta - g\\
   \ddot{\theta} &= \tau
\end{split}
\end{equation}

where, $\tau = \frac{rF_1}{J}$, $f=\frac{F_2}{m}$, and $\epsilon = \frac{J}{mr}$. System \eqref{eq:PVTOL} reduces to the one in \cite{Sastry1991} in the normalized units $g=1$. Here $\epsilon$ is a small positive constant, which denotes the coupling factor present between rolling and lateral motion.

\subsection{Feedback Linearization}
Consider the output position $(x,y)$ and differentiating it twice yields
\begin{equation}\label{eq:FBL01}
\begin{bmatrix} \ddot{x} \\ \ddot{y} \end{bmatrix} = \begin{bmatrix} 0 \\ -g \end{bmatrix} + \begin{bmatrix} -\sin\theta & \epsilon\cos\theta \\ \cos\theta & \epsilon\sin\theta \end{bmatrix} \begin{bmatrix} f \\ \tau \end{bmatrix}
\end{equation}
since the decoupling matrix is non-singular for $\epsilon > 0$, though barely, Eq. \eqref{eq:PVTOL} has a vector relative degree of $[2, 2]$. Thus exact we get the following control law using feedback linearization.
\begin{equation}\label{eq:FBL02}
\begin{bmatrix} f \\ \tau \end{bmatrix} = \begin{bmatrix} -\sin\theta & \cos\theta \\ \frac{1}{\epsilon}\cos\theta & \frac{1}{\epsilon}\sin\theta \end{bmatrix} \left( \begin{bmatrix} 0 \\ g \end{bmatrix} + \begin{bmatrix} v_1 \\ v_2 \end{bmatrix} \right)
\end{equation}
where, $v_1$ and $v_2$ are virtual inputs, and results in following transformed system,
\begin{equation}\label{eq:FBL03}
\begin{split}
   \ddot{x}    &= v_1 \\
   \ddot{y}    &= v_2 \\
   \ddot{\theta} &= \frac{1}{\epsilon}\left[v_1 \cos\theta + (v_2 + g)\sin\theta\right]
\end{split}
\end{equation}
Considering the zero dynamics of the above system, by restricting the outputs, their derivatives, and virtual inputs to zero. This gives the following unstable zero dynamics.
\begin{equation}\label{eq:FBL04}
  \ddot{\theta} = \frac{g}{\epsilon} \sin\theta
\end{equation}
This clearly shows that the PVTOL  \eqref{eq:PVTOL} is a non-minimum phase system, and therefore by using feedback linearization, we cannot achieve internal stability.

\section{Inverse Optimal Control Design}\label{sec:InvOPT}
As discussed in section \ref{sec:Prelim} a CLF needs to be constructed for inverse optimal control design. However, as shown in the previous section, PVTOL dynamics cannot be linearized by a static feedback law, due to unstable zero dynamics. So we will use a dynamic feedback linearization approach to derive the normal form and thus construct the CLF. Using the well known transformation proposed by \cite{OlfatiSaber2002} as follows,
\begin{equation}\label{eq:AFBL01}
\begin{split}
   \hat{x}      &\triangleq x - \epsilon\sin\theta \\
   \hat{y}      &\triangleq y + \epsilon(\cos\theta - 1)
\end{split}
\end{equation}

It is worth noting that the above transformation is diffeomorphic and maps the equilibrium point $(x_0,y_0,0)$ to $(\hat{x}_0,\hat{y}_0,0)$. Applying the transformation \eqref{eq:AFBL01} to \eqref{eq:PVTOL} yields,
\begin{equation}\label{eq:AFBL02}
\begin{split}
   \ddot{\hat{x}} &= - \hat{f} \sin\theta \\
   \ddot{\hat{y}} &= \hat{f} \cos\theta - g\\
   \ddot{\theta}  &= \tau
\end{split}
\end{equation}
where $\hat{f}\triangleq f - \epsilon \dot{\theta}^2$. Considering the transformed position $(\hat{x},\hat{y})$ as output, and differentiating it twice yields
\begin{equation}\label{eq:AFBL03}
\begin{bmatrix} \ddot{x} \\ \ddot{y} \end{bmatrix} = \begin{bmatrix} 0 \\ -g \end{bmatrix} + \begin{bmatrix} -\sin\theta & 0 \\ \cos\theta & 0 \end{bmatrix} \begin{bmatrix} \hat{f} \\ \tau \end{bmatrix}
\end{equation}

Now since the decoupling matrix is singular no static feedback linearization law exists. But since $\ddot{\theta}=\tau$, so differentiating Eq. \eqref{eq:AFBL03} two more times gives,
\begin{equation}\label{eq:AFBL04}
\begin{bmatrix} \hat{x}^{(4)} \\ \hat{y}^{(4)} \end{bmatrix} = \phi + G  \hat{u}
\end{equation}

where $\hat{u} = \begin{bmatrix} \ddot{\hat{f}} & \tau \end{bmatrix}^\top$, and
\[
\phi \triangleq \begin{bmatrix} \hat{f}\dot{\theta}^2 \sin\theta - 2\dot{\hat{f}} \dot{\theta} \cos\theta \\ - \hat{f}\dot{\theta}^2 \cos\theta - 2\dot{\hat{f}} \dot{\theta} \sin\theta \end{bmatrix}, \quad
G \triangleq \begin{bmatrix} -\sin\theta & -\hat{f}\cos\theta \\ \cos\theta & -\hat{f}\sin\theta \end{bmatrix}
\]

So, the dynamic feedback law, which linearizes the transformed system \eqref{eq:AFBL02} will be,
\begin{equation}\label{eq:AFBL05}
\hat{u}_{FL} = G^{-1}\left(-\phi + v\right)
\end{equation}
where $v$ is the virtual control input to the linearized system.

\begin{remark}
It must be noted the linearized system (system \eqref{eq:AFBL04} with control law \eqref{eq:AFBL05}) does not contain any unobservable (zero) dynamics \cite{Sastry1991}.
\end{remark}

Let $z = [\hat{x},\dot{\hat{x}},\ddot{\hat{x}},\dddot{\hat{x}},\hat{y},\dot{\hat{y}},\ddot{\hat{y}},\dddot{\hat{y}}]^T$.
The system \eqref{eq:AFBL04} can be written in the following state space form,

\begin{equation}\label{eq:IPC01}
\dot{z} = \Phi + \Gamma\hat{u}
\end{equation}
where,
\[
\Phi = \begin{bmatrix}
         z_2 \\
         z_3 \\
         z_4 \\
         \hat{f}\dot{\theta}^2 \sin\theta - 2\dot{\hat{f}} \dot{\theta} \cos\theta \\
         z_6 \\
         z_7 \\
         z_8 \\
         - \hat{f}\dot{\theta}^2 \cos\theta - 2\dot{\hat{f}} \dot{\theta} \sin\theta
       \end{bmatrix},\quad
\Gamma = \begin{bmatrix}
           0 & 0 \\
           0 & 0 \\
           0 & 0 \\
           -\sin\theta & -\hat{f}\cos\theta \\
           0 & 0 \\
           0 & 0 \\
           0 & 0 \\
           \cos\theta & -\hat{f}\sin\theta
         \end{bmatrix}
\]

Since system \eqref{eq:IPC01} is in normal form and can be transformed into a linearized system by feedback law \eqref{eq:AFBL05}, a control Lyapunov function exists for system \eqref{eq:IPC01} of the form
\begin{equation}\label{eq:IPC02}
 V = \frac{1}{2}z^TPz
\end{equation}

Here $P$, a positive definite matrix, can be selected to satisfy the following Ricatti inequality

\begin{equation}\label{eq:IPC03}
A^TP + PA - PBB^TP < 0
\end{equation}

where $A$ and $B$ are state-space realization of linearized system ($\dot{z}=Az+Bv$) and can be written as follows,
\[
A = I_2 \otimes \begin{bmatrix}
                  0 & 1 & 0 & 0 \\
                  0 & 0 & 1 & 0 \\
                  0 & 0 & 0 & 1 \\
                  0 & 0 & 0 & 0
                \end{bmatrix},\qquad
B = I_2 \otimes \begin{bmatrix} 0 \\ 0 \\  0 \\ 1 \end{bmatrix}
\]

where $I_2$ is $2\times2$ identity matrix, and $\otimes$ represents kronecker product. Due to the structure of $A$ and $B$ matrices, Lyapunov matrix $P$ can be selected of the following form,
\begin{equation}\label{eq:IPC04}
  P = \begin{bmatrix}
        k_x & 0 \\
        0 & k_y
      \end{bmatrix} \otimes P_0
\end{equation}
where $k_x,k_y$ are tunable gains, and $P_0$ is a constant matrix shown below, obtained using solving corresponding LMI using YALMIP \cite{Lofberg2004} and SeDuMi \cite{Sturm1999}.
\[
P_0 = \begin{bmatrix}
0.25 & 0.40 & 0.95 & 0.70\\
0.40 & 2.40 & 4.00 & 3.80\\
0.95 & 4.00 & 9.80 & 9.40\\
0.70 & 3.80 & 9.40 & 13.0
\end{bmatrix}
\]
and it can be shown that for all $k_x, k_y \in [0.2, 10^6]$, $P$ is positive definite and satisfies the Ricatti  inequality \eqref{eq:IPC03},

Now since we have the CLF, Sontag’s formula \eqref{eq:Prelim07} can be used to achieve given inverse optimal control law for \eqref{eq:IPC01},
\begin{equation}\label{eq:IPC05}
\hat{u}_{InvOpt} = \begin{cases} - (c_0 + \lambda)\Gamma^\top Pz, &\quad \Gamma^\top Pz \ne 0 \\
         \hfil 0,  &\quad \Gamma^\top Pz = 0
   \end{cases}
\end{equation}
where,
\[
\lambda = \dfrac{z^\top P \Phi + \sqrt{(z^\top P \Phi)^2 + (z^\top P\Gamma\Gamma^\top Pz)^2}}{z^\top P\Gamma\Gamma^\top Pz}
\]

The inverse optimal controller \eqref{eq:IPC05}, by Theorem \ref{thm:02}, minimizes the following cost functional and have sector margin of $(\frac{1}{2},\infty)$ by Proposition \ref{prop:01} and \ref{prop:02}.
\begin{equation}\label{eq:IPC06}
J = \int_{0}^{\infty} \left[ \frac{\mu}{2}z^\top P\Gamma\Gamma^\top Pz + \frac{1}{2\mu}\hat{u}^\top \hat{u} \right] \,\mathrm{d}t
\end{equation}
where,
\[
\mu = \begin{cases} c_0 + \lambda, &\quad \Gamma^\top Pz \ne 0 \\
         \hfil c_0,  &\quad \Gamma^\top Pz = 0
   \end{cases}
\]

\section{Results and Discussion}\label{sec:Res}
\begin{figure}
  \centering
  \includegraphics[width=\linewidth]{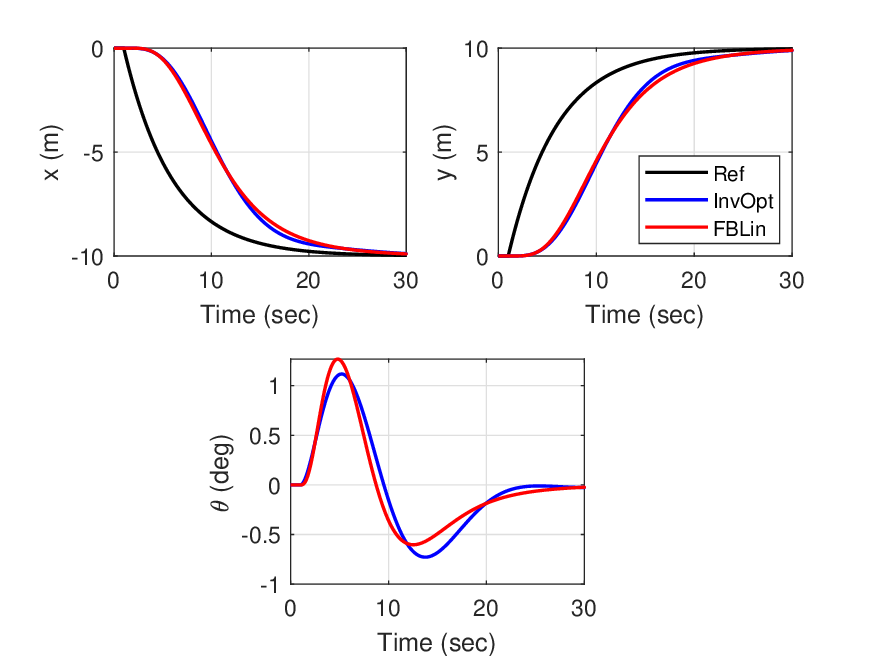}
  \caption{Simulation Results: Position and Attitude}
  \label{fig:SimResPos}
\end{figure}
\begin{figure}
  \centering
  \includegraphics[width=\linewidth]{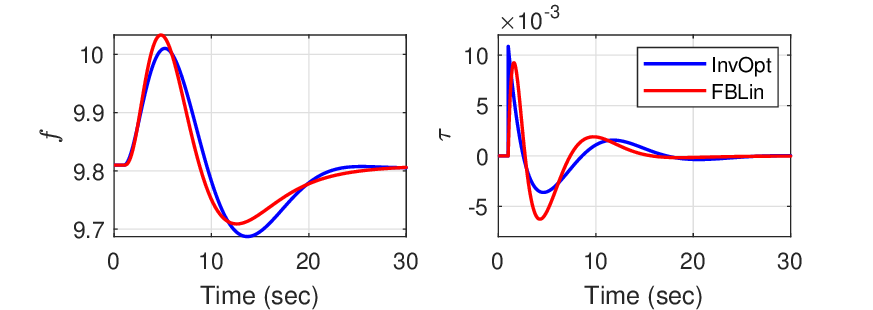}
  \caption{Simulation Results: Control Effort}
  \label{fig:SimResContEff}
\end{figure}
\begin{figure}
  \centering
  \includegraphics[width=0.75\linewidth]{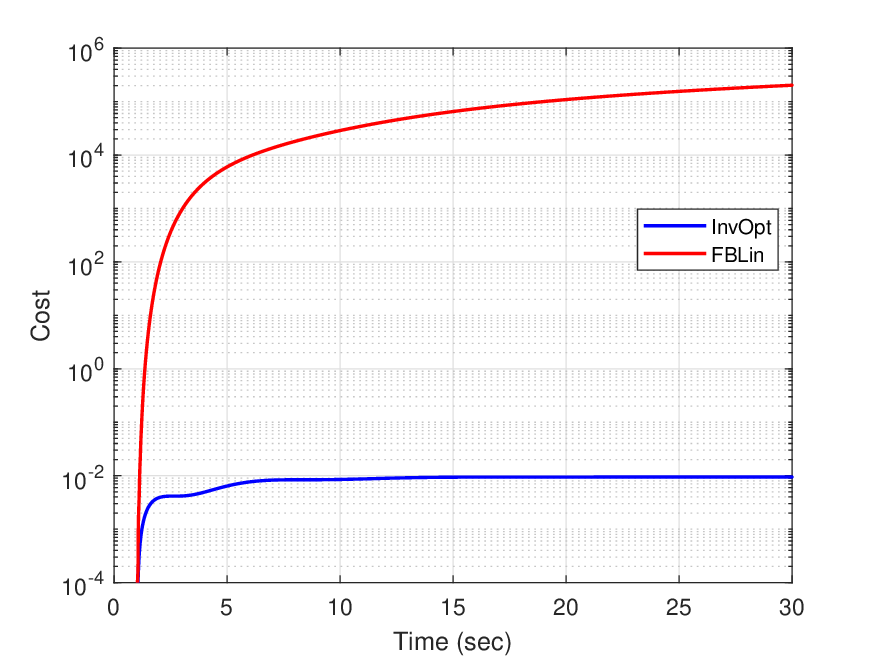}
  \caption{Simulation Results: Cost Function}
  \label{fig:SimResCst}
\end{figure}

To demonstrate the effectiveness of inverse optimal control law \eqref{eq:IPC05}, nonlinear simulation results are presented in this section. Results are compared with that of dynamic feedback linearization based control law \eqref{eq:AFBL05}. For the inverse optimal controller, tunable gains $c_0,k_x,k_y$ are all equal to 1. Moreover, the virtual control input ($v$) of \eqref{eq:AFBL05} is selected to be state feedback ($v = -Kz$), where $K=I_2\otimes K_0$, and  for a fair comparison $K_0 = [0.1875, 1.0375, 2.4, 2.55]$ is selected using pole placement approach to achieve the closed loop response similar to that of inverse optimal controller. For both controllers' simulations, very strong coupling ($\epsilon = 1$) is used and for acceleration due to gravity ($g$) value of 9.81 is used. It is interesting to mention that both proposed controllers are quite robust towards the coupling parameter ($\epsilon$), mainly due to the transformation used in their design.

Figures \ref{fig:SimResPos} and \ref{fig:SimResContEff} show the position tracking, attitude response, and control efforts, respectively. Figure \ref{fig:SimResCst} shows the cumulative cost, defined in \eqref{eq:RES01}, it can be easily seen that for up to 30 seconds, the cost of inverse optimal controller \eqref{eq:IPC05} is about $0.0094$ while that of feedback linearization based control law \eqref{eq:AFBL05} is about $2\times 10^5$, despite the fact that the linear controller gains of \eqref{eq:AFBL05} are selected to have  a similar performance with both controllers.
\begin{equation}\label{eq:RES01}
\hat{J}(t) = \int_{0}^{t} \left[ \frac{\mu}{2}z^\top P\Gamma\Gamma^\top Pz + \frac{1}{2\mu}\hat{u}^\top \hat{u} \right] \,\mathrm{d}\tau
\end{equation}

\begin{figure}
  \centering
  \includegraphics[width=\linewidth]{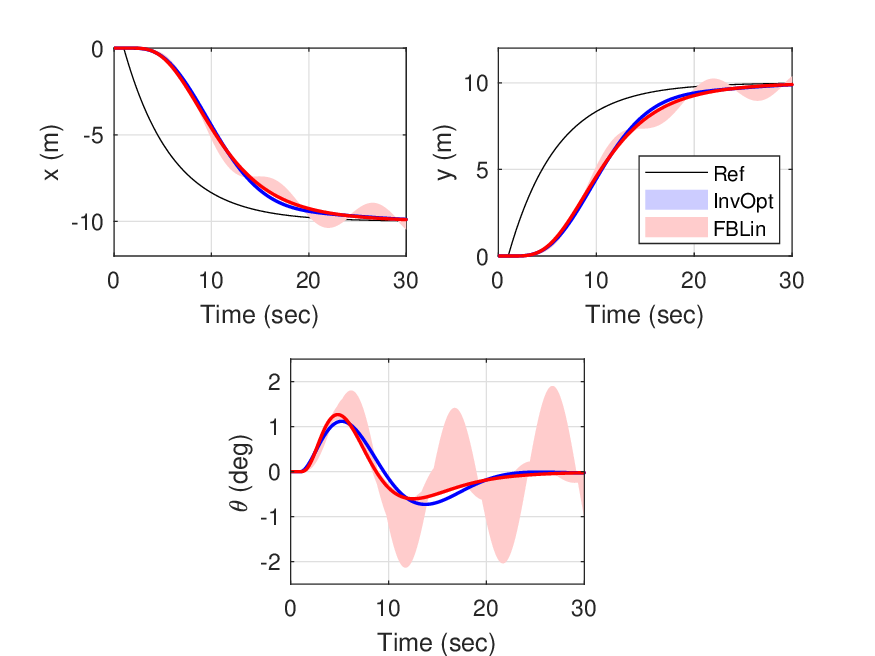}
  \caption{Monte-Carlo Simulation Results: Position and Attitude}
  \label{fig:SimResPosPerturbed}
\end{figure}

\begin{figure}
  \centering
  \includegraphics[width=\linewidth]{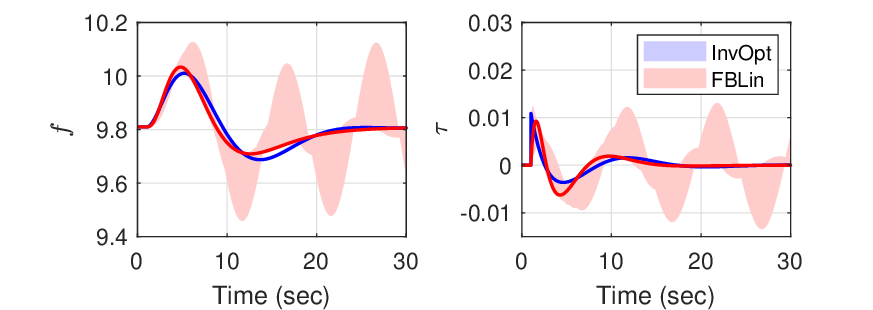}
  \caption{Monte-Carlo Simulation Results: Control Effort}
  \label{fig:SimResContEffPerturbed}
\end{figure}

To demonstrate the inverse optimal control law's robustness, the perturbed input $\Delta \hat{u}$ is applied to system \eqref{eq:IPC01} for both feedback linearization \eqref{eq:AFBL05} and inverse optimal based control law \eqref{eq:IPC05}, where  $\Delta = \mathrm{diag}([\delta_f,\delta_\tau])$. Monte-carlo type simulations are performed with $\delta_f,\delta_\tau \in [0.2,5]$, to assess the MIMO gain margins at input of  \eqref{eq:IPC01}, and results are shown in figures \ref{fig:SimResPosPerturbed} and \ref{fig:SimResContEffPerturbed}, where shaded regions shows the responses in uncertain cases, while the solid lines represents the nominal scenario. It must be noted that both uncertain gains ($\delta_f,\delta_\tau$) are varied independently and simultaneously in 100 simulation runs. It can be seen that with feedback linearization based control law, the closed-loop is almost on the verge of instability as depicted by the sustained oscillations in attitude ($\theta$) and control efforts, while oscillations in position are small. However, with inverse optimal control law, results show excellent robustness characteristics without any noticeable degradation of performance, despite such large variations in uncertain gains.

\section{Conclusion}\label{sec:Conc}
The paper presents a constructive approach to design an inverse optimal nonlinear controller for highly coupled and non-minimum phase VTOL aircraft, using Control Lyapunov Function. Later, a comprehensive comparison is made between an inverse optimal controller and a dynamic feedback linearization based controller. The proposed controller, being optimal, possesses some excellent robustness properties. It was demonstrated through simulation results that the inverse optimal controller achieves outstanding robustness with minimum performance degradation against a perturbed input despite large gain variations.

\bibliography{References}
\end{document}